\documentstyle[prl,aps]{revtex}
\begin{document}
\title
{Maximum Accretion Efficiency in General Theory of Relativity}

\author{Abhas Mitra}
\address{Theoretical Physics Division, Bhabha Atomic Research Center,\\
Mumbai-400085, India\\ E-mail: amitra@apsara.barc.ernet.in}


\maketitle

\begin{abstract}
We derive here the expression for the accretion luminosity, $L(\infty)$,
as seen by a distant inertial observer $S_\infty$, for the case of
spherical accretion onto a static compact object having a surface
gravitational red-shift $z_x$.  It is found that the ``efficiency'' for
conversion of mass energy into accretion energy is given by $\epsilon =
z_x/(1+z_x)$.  And since the maximum value of $z_x$ permitted by General
Theory of Relativity (GTR) is  2, the maximum theoretical value of the
accretion efficiency is 66.66$\%$.

\end{abstract}


\section{Introduction}
One of the key concepts in astrophysics is that of accretion of the ambient 
gas by a
compact object by virtue of its gravitational field. In the test particle assumption,
as the accreting
material traverses down the gravitational potential well of the compact
object having a mass $M_x$ and a {\em hard surface} of radius $R_x$, the infalling kinetic
energy is released when the gas ``hits'' the ``hard surface''. The rate of
 energy  release by this process is given by the depth of the potential well :
and, for an accretion rate $\dot M$, the accretion luminosity is
given by [1]
\begin{equation}
L = {G M_x {\dot M} \over R_x}
\end{equation}
where $G$ is the gravitation constant.
When the gas possesses viscosity, it gets heated during the infall unlike
an idealized test particle or a dust, and radiates even before it heats the
``hard surface''.  In this case, the kinetic energy released upon impact
is less than the corresponding free fall value. Accordingly,
 the amount of accretion 
energy released after the
impact is less than the corresponding value for a dust. However, the net
energy released over the entire process is same in either case.
Thus, viscosity or no viscosity, the eventual expression for $L$
remains unchanged.

A related question is the idea of the maximum luninosity or the Eddington
luminosity at which the repulsive effect of the emitted radiation cancels
the gravitational attraction on the infalling particles [1]: 
\begin{equation}
L_{ed} = {4 \pi G M_x c\over \kappa}
\end{equation}
where $c$ is the speed of light and $\kappa$ is the appropriate opacity of the accreting plasma.
If we are considering emission of electromagnetic radiation only, for a 
fully ionized
H-plasma, the expression for opacity is given by $\kappa = \sigma_T/m_p$,
where $\sigma_T= 6.65 \times 10^{-25}$cm$^2$ is the Thompson cross section
and $m_p$ is the proton mass , so that $\kappa \approx 0.4$ cm$^2$/g. The corresponding
Eddington luminosity is given by:
\begin{equation}
L_{ed} = 1.26 \times 10^{38}  ~\left({M_x\over M_\odot}\right) ~ergs/s
\end{equation}
However, at very high accretion rates, astrophysical plasma may cool by
emitting neutrinos, and in such cases, the value of $L_{ed}$ could be much
higher because of the tiny neutrino-matter intraction cross-sections [1]:
\begin{equation}
L_{ed} \approx 2 \times 10^{54} ~ \left( {T_\nu \over 8 MeV}\right)^{-2} ~ergs/s ~\left({M_x\over M_\odot}\right)
\end{equation}
where $T_\nu$ is the temperature of the emitted neutrinos.

It is obvious that all the above expressions are correct only for the
Newtonian gravity and the aim of this paper is to arrive at the correct
General Relativistic (GTR) expressions for the above quantities.

\section{Energy and Luminosity in GTR}
In GTR, the global notions like the total mass-energy of an isolated body
$M$, its self-gravitational energy, $E_g$, and the binding energy, $E_B$, are
meaningfully definable only with respect to an observer $S_\infty$
situated at infinite spatial distance. This is so because, the spacetime
seen by him is flat (Galilean) and where the notions of global energy-momentum
 conservation can
be readily invoked. At the same time it is known that the energy of a
particle/photon emitted from the surface of the compact object is reckoned
to be lower by a factor of ($1+z$) by $S_\infty$. Consequently, if the
temperature/energy of  the particles/photons near the surface of the compact
object is $T$, its value as measured by $S_\infty$ is [1,2,3,4]
\begin{equation}
T_\infty = T (1+z)^{-1}; \qquad T = T_\infty (1+z)
\end{equation}
Also, since the clocks move slower near the compact object by the same factor,
the local value of luminosity is higher by a factor of $(1+z)^2$:
\begin{equation}
L = L({\infty})  (1+z)^2; \qquad L({\infty}) = L (1+z)^{-2}
\end{equation}
This fact is well known and yet we discussed it here for the sake of
completion. Accordingly, the fact that, the expression for Eddington luminosity
gets modified in GTR is well discussed in literature [1]:
\begin{equation}
L_{ed} = {4 \pi G M_x c (1+z)\over \kappa}
\end{equation}
and
\begin{equation}
L_{ed}({\infty}) = {4 \pi G M_x c \over \kappa (1+z)}
\end{equation}
Here, note that, the definition of $E_{ed}$ does not depend on the actual
definition of $L$, and the GTR correction arises from the fact that in an
external Schwarzschild spacetime, the definition of gravitational
acceleration gets modified. It is also pointed out that the notion of
Eddington luminosity is relevant only locally, i.e, at the source of the
radiation. Thus, it is actually, the Eq. (7) which is relevant here.
However, surprisingly, what has not been discussed in literature is how the
basic expression for {\em accretion luminosity} should change in GTR
(although, we will point out later that the expression for the binding energy
of a rotating pressureless infinetisimally thin disk is known).
We shall obtain this GTR expression for accretion luminosity in a remarkably
simple manner.

\section{Accretion Luminosity}
Since, in GTR, the concept of the mass of an isolated body (for instance
$M_x$) is defined with respect to $S_\infty$, it is necessary that the
accretion rate is measured by the same distant observer:
\begin{equation}
{\dot M}_{\infty} \sim {\Delta M \over \Delta t}\mid_\infty
\end{equation}
Accretion energy arises from the conversion of infall kinetic energy into
random energy. If a hard surface is present, the entire available kinetic
energy is converted into random energy, and, on the other hand, if at a
given region $R >R_x$, where no hard surface is present, the gas can still
convert part of its infall kinetic energy into random energy and even
radiate it. In the limit of infinite viscosity, the energy emitted at $R$
could be equal to the infall kinetic energy, and, thus, the {\em maximum}
accretion energy at any region, $R \ge R_x$, is determined by the  free fall
kinetic energy. If $v$ is the speed of the test particle measured in a
Local Inertial Frame (LIF), in, Newtonian physics
\begin{equation}
Kinetic ~Energy ~Per ~Unit ~Mass = {1\over 2} v^2
\end{equation}
Also, for free-fall, one has
\begin{equation}
{1\over 2} v^2 = {G M_x \over R}
\end{equation}
And, as is well known, the two foregoing equations trivially explain the
origin of the Newtonian formula for accretion luminosity (Eq. 1).

For a transition to GTR, first we recall that, by the Principle of Equivalence,
GTR reduces to the Special Theory of Relativity in the LIF. So, in the
LIF, we have [1,2,3,4]:
\begin{equation}
Kinetic ~Energy ~Per ~Unit ~Mass = \left[\left(1- {v^2\over c^2}\right)^{-1/2}
-1\right] c^2
 =\left(\gamma -1\right)c^2
\end{equation}
where $\gamma$ is the Lorentz factor of the fluid. However, in an external Schwarzschild geometry, in the absence of angular momentum,
the equation (11) remains exactly valid [1,2,3,4], so that
\begin{equation}
Kinetic ~Energy ~Per~ Unit ~Mass = \left[\left(1 - {2 G M_x\over R
c^2}\right)^{-1/2} -1\right] c^2
\end{equation}
The right hand side of the above equation can be identified with the
gravitational redshift for photons (or any paticles) emitted at $R=R$:
\begin{equation}
Kinetic ~Energy ~Per ~Unit ~Mass = (\gamma -1) c^2 = z c^2
\end{equation} 
It is also clear that $zc^2$ is the escape kinetic energy or gravitational
binding energy of a test particle at $R=R$.
As discussed before, because of gravitational red shift, energy released
at $R=R$ will appear lower to $S_\infty$ by a factor $(1+z)$.
Therefore, the maximum accretion energy per unit mass or, the accretion
efficiency seen by $S_\infty$ will be
\begin{equation}
\epsilon = {z_x \over 1+z_x}
\end{equation}
And hence the accretion luminosity measured by $S_\infty$ is
\begin{equation}
L (\infty) = {z_x \over 1+z_x} {\dot M}_\infty c^2
\end{equation}
Eq.(16) obviously reduces to the Newtonian Eq.(1) for $GM_x/R_x c^2 \ll 1$.
Also, the accretion luminosity, measured locally will be higher by a
factor $(1+z)^2$:
\begin{equation}
L = z_x (1+z_x) {\dot M}_\infty c^2
\end{equation}

Now, we would like to point out an early work which handled the more
difficult problem of finding the gravitational binding energy of an
idealized rotating infinetisimally thin disk of a dust (pressure =0) [5]:
\begin{equation}
E_B = M_0 c^2 \left[ {z_c \over 1+z_c} - {2 \omega J\over M_0 c^2}\right]
\end{equation}
where $M_0$ is the ``rest mass'' (measured by $S_\infty$ of course),
$\omega$ is the angular speed, and $J$ is the angular momentum of the
disk. $z_c$ is the redshift at the ``center'' of the infinite disk. Here
the binding energy of the disk (as measured by $S_\infty$) is given by:
\begin{equation}
E_B = (M_0 -M) c^2
\end{equation}
In the absence of rotation, the Eq.(18) would give $E_B = {z_c\over 1+z_c} M_0
c^2$. And although, the problem of spherical accretion discussed by us is
technically different from the above problem, clearly, there is an inner
physical link and agreement.
\section{Discussion}
GTR gives an absolute upper limit on the compactness of cold bodies in
hydrostatic equilibrium. This limit is independent of the details of the
equation of state of the stellar matter, and is given by [2,3,4]
\begin{equation}
{2G M_x\over R_x c^2} \le {8\over 9}; \qquad z_x \le 2; \qquad \gamma_x
\le 3
\end{equation}
Thus, the theoretical maximum accretion efficiency is $\approx 66.66 \%$.
We recall that, the accretion efficiency for a canonical neutron star is
$\sim 10 -15 \%$. In contrast, for strict spherical accretion onto a
Schwarzschild black hole, no energy is emitted when the accreted material
hits the ``event horizon'' because event horizon is not a ``hard surface''
and is only a fictitious membrane with $z=\infty$. In this case accretion
luminosity, would be only due to viscous and radiative properties of the infalling gas, and
could be very low. This specific feature is invoked for the observational
search for BHs. However, in astrophysics, accretion matter usually
possesses angular momentum, and accretion is usually mediated by
an ``accretion disk''. For thin and sufficiently viscous accretion disks
around Schwarzschild BHs, the
``inner edge'' is located at three Schwarzschild radius: $R_i = 6 G
M_x/c^2$. It is found that, the maximum accretion efficiency in such a
case is $\epsilon \approx 5.72 \%$ [1,2], and the accretion energy output comes
in the form of photons or neutrinos because there is no emission of
gravitational radiation in a stationary gravitational field.

On the other hand, the net maximum efficiency for BH accretion could be
much higher when the BH is rotating, or, in other words, if we have a Kerr
BH. The maximum value of $\epsilon$ is realized for a maximally rotating
BH,i.e., for which one has $J= G M^2/c$, and is given by $\epsilon \approx
42.3\%$ [1,2]. Further, the accreting matter in the rotating Kerr spacetime may
emit radiation by means of gravitational waves, and, it can not be
ascertained, what fraction of this efficiency corresponds to emission of
photons and neutrinos. In any case, the value of $\epsilon \approx 66.66 \%$ obtainable for a
compact object with $z_x =2$ is indeed the corresponds to the highest
theoretical accretion efficiency.

Now, the natural question would be whether, this high accretion efficiency
can be realized in Nature. The canonical NS is a hard
-surface compact
object with a modest value of $  0.1 < z_x<0.2$. And GTR actually allows
existence of static compact objects upto $z_x =2$. If in future, there is
much improved
undrstanding of nuclear equation of state (EOS) at extremely high temperatures
and densities, and further, if it is possible to solve the GTR collapse
equations with full generality and without making tacit simplification, it
might be possible to theoretically infer the existence of Ultra Compact
Objects even in the context of normal QCD. Note that, a total accumulated
theoretical and numerical uncertainty of $\sim 10\%$ can push the
 $2GM/R c^2 =8/9$ state ($z_x =2$) to an
apparently event horizon state with $2GM/R c^2 =1$ ($z = \infty$). Thus, in
the absence of an exact EOS and radiation transport ansatz at arbitrary
high density and temperature, the reliability
of any numerical (approximate) study
 of the GTR collapse problem is rather poor once we proceed 
beyond the NS stage.

It may be pointed out that some effective field theories
of strong interaction allow confinement of not only quarks but also of
nucleons at densities  {\em well below} the nuclear  (rest mass)
density, $\rho_{nu}
\approx 2.8 \times 10^{14} $ g cm$^{-3}$ [6,7,8]. Compact objects proposed on this
idea are called Q-stars (Q stands for a conserved quantum number and nor
for ``quark''). Q-stars could be very massive $\ge 100 M_\odot$. For some
choice of the model parameters, the maximum value of $2GM/Rc^2 <1/3$ [9]
corresponding to $z_x \le 0.73$. Clearly, since Q-stars may have densities
below the nuclear densities, depending on the model parameters, it may
also be possible to have such stars with $z <0.1$, stars less compact than
a canonical NS.
Presumably, a different EOS may lead to
higher values of $z_x$. In other words, it is possible that, beyond the
White dwarf stage, Nature actually allows existence of compact objects
having a fairly wide range of $z_x$ with an upper bound at $z_x =2.0$. In fact, at present, there is no proper
explanation for the cosmic gamma ray bursts which may involve release of
$\sim 10^{55}$ergs of neutrinos [10]. It is presumable that the gamma ray
bursts are birth pangs of relativistic ultra compact objects.


\end{document}